\documentclass[a4paper,10pt,twocolumn]{article} 
\usepackage{graphicx}
\usepackage{subfigure}
\usepackage{amsmath}
\usepackage{amssymb}
\usepackage{wrapfig}
\usepackage{color}
\usepackage[english]{babel}
\usepackage[warning,math]{easyeqn}
\usepackage[thinlines]{easybmat}
\usepackage{enumerate} 
\usepackage{multirow} 
\usepackage[pdftex,
            pdfauthor={Yves-Henri Sanejouand},
            pdftitle={Normal-mode driven exploration of protein domain motions},
            pdfsubject={Generating protein conformers},
            pdfkeywords={Normal modes, Conformers, Distance restraints, Rosetta},
            pdfproducer={Latex with hyperref},
            pdfcreator={Texmaker}]{hyperref}
\begin{document}

\title{\textbf{Normal-mode driven exploration of \\ 
protein domain motions}}

\author{Yves-Henri Sanejouand\footnote{yves-henri.sanejouand@univ-nantes.fr}\\ \\
		UFIP, UMR 6286 of CNRS,\\
        Facult\'e des Sciences et des Techniques, Nantes, France.} 
\date{March 22$^{th}$, 2021}
\maketitle

\section*{Abstract}

Domain motions involved in the function of proteins can often be well described as a combination of motions along a handfull of low-frequency modes, that is, with the values of a few normal coordinates.
This means that, when the functional motion of a protein is unknown, it should prove possible to predict it, since it amounts to guess a few values. 

However, without the help of additional experimental data, using normal coordinates for generating accurate conformers far away from the initial one is not so straightforward.
To do so, a new approach is proposed: 
instead of building conformers directly with the values of a subset of normal coordinates, 
they are built in two steps, the conformer built with normal coordinates being just used for defining a set of distance constraints, the final conformer being built so as to match them.
Note that this approach amounts to transform the problem of generating accurate protein conformers using normal coordinates into a better known one: the distance-geometry problem, which is herein solved with the help of the ROSETTA software.

In the present study, this approach allowed to rebuild accurately six large amplitude conformational changes, using at most six low-frequency normal coordinates. As a consequence of the low-dimensionality of the corresponding subspace, random exploration also proved enough for generating low-energy conformers close to the known end-point of the conformational change of the LAO binding protein, lysozyme T4 and adenylate kinase. 

\vskip 1cm
\textbf{Keywords}: Conformers, Normal modes, Distance constraints, Elastic Network Model, ROSETTA.

\section*{Introduction}

It has been known for a while that domain motions involved in the function of a protein can be well described as a combination of motions along its lowest-frequency normal modes \cite{Harrison:84,Brooks:85,Marques:95,Perahia:95}, and this even when the normal modes are obtained using simplified protein models \cite{Harrison:84,Tama:00,Tama:01,Delarue:02,Gerstein:02,Mahajan:15}, like the elastic network model (ENM) \cite{Tirion:96,Bahar:97,Hinsen:98}.  

These results suggest that it should prove possible to predict the functional motion of a protein, since it amounts to guess the values of a few normal coordinates. 
Indeed, relevant conformers can be obtained by optimizing these values so as to match low-resolution X-ray crystallographic \cite{Tirion:95,Ma:10}, Cryo\-EM \cite{Tama:03,Norma,Delarue:04} or SAXS \cite{Tama:10,Svergun:16} data, to solve molecular replacement problems \cite{Delarue:04,Elnemo1} or to fulfill a set of experimentally determined distance constraints \cite{Brooks:05}.

On the other hand, when additional experimental data are missing, normal mode coordinates can be used in order to explore the conformational landscape of a protein in the vicinity of a known structure, for instance through single mode-following techniques \cite{Derreumaux:02,Jordan:06,Perahia:06,Robert:11}.
They can also be used so as to take advantage of the low-dimensionality of the subspace to be explored, in conjunction with various sampling methods \cite{Bahar:20}, such as random exploration \cite{Ritchie:12}, energy minimization \cite{Abagyan:05,Zacharias:08,Bates:10,Doruker:16}, Monte Carlo Metropolis \cite{Go:85,Jordan:07}, molecular dynamics \cite{Durup:91,Elezgaray:98,Perahia:15}, path planning protocols \cite{Cortes:08} or geometric simulations \cite{Wells:12}.      

In principle, they could also be used to perform straight jumps, that is, to go from a conformer to another one in a single step, noteworthy in the case of pairs of conformers with a C$_\alpha$-root-mean-square difference (RMSd) of several {\AA}ngstr\"oms. 

Indeed, it was shown in a previous study that, with the help of algorithms implemented in the ROSETTA software \cite{Rosetta:15}, it is possible to perform such jumps. Noteworthy, in the case of 14 large amplitude (RMSd $>$ 2 {\AA}) conformational changes, it proved possible to obtain conformers less than 1 {\AA} away from the target, that is, the known end-point of the conformational change \cite{Mahajan:17}. 
Though encouraging, such successes proved rare, namely, 15\% of the conformational changes considered, underlining the fact that using normal coordinates for generating accurate conformers far away from the initial one is not a straightforward task \cite{Sharp:09,Gohlke:11,Lavery:19}.

In the present study, a promising solution is proposed. 
Instead of building conformers using directly a set of normal coordinates, they are built in two steps, the conformer built with normal coordinates being just used for defining a large set of distance constraints, the final conformer being built so as to match them.
Note that this approach amounts to transform the problem of generating accurate protein conformers using normal coordinates into a better known one: the distance-geometry problem \cite{Wuthrich:87,Zwu:99,Malliavin:19}.  

\section*{Methods}

\subsection*{Building conformers using normal coordinates}

Conformers can be generated by displacing the $N$ atoms of a system along $n$ of its normal modes ($n \leq 3N$), $\Delta r_i$, the displacement of coordinate $i$ being:
\begin{equation}
\Delta r_i = \frac{1}{\sqrt{m_i}} \sum^n_{k=1} a_{ik} q_k
\label{eq:dr}
\end{equation}
where $q_k$ is the k$^{th}$ normal coordinate of the system, that is, the amplitude of the motion along mode $k$, $a_{ik}$ being the weight of mode $k$ for coordinate $i$, while $m_i$ is the mass of the corresponding atom.

In the present study, the elastic energy (eqn \ref{eq:enm}) of the conformers obtained using eqn \ref{eq:dr} is usually minimized, under the following set of constraints:
\[
q_k = cste, \hspace{1cm} k=1 \dots n 
\]
In other words, during energy minimization, only the $3N -n$ other coordinates are allowed to vary, as done in a number of previous studies \cite{Jordan:06,Perahia:06,Mahajan:17}. 

In practice, conformer generation was performed with the dq\_modes software, which is freely available on HAL (\url{https://hal.archives-ouvertes.fr/hal-03144518}), via the Software Heritage program \cite{Softwareheritage}. 

The conformers thus obtained were not considered as is. Instead, they were used to specify NOE-like restraints between all pairs of C$_\alpha$ atoms of the protein considered, except for pairs of C$_\alpha$ atoms belonging to consecutive residues along the sequence, the corresponding C$_\alpha$-C$_\alpha$ distances being not expected to vary significantly during a conformational change.

Conformers consistent with this set of distance restraints were then determined with the ROSETTA software \cite{Rosetta:15}, version 2018.12, using the RELAX protocol, a 1 {\AA} distance between the lower and upper NOE boundaries being assumed in order to take into account the lack of precision of the distances obtained above. Finally, all restraints were removed and an energy minimization was performed, with the parameters of the MINIMIZE protocol set to their default values.

Note that hereafter, for the sake of comparison between proteins of different sizes, minimum energies thus obtained are given per residue.  

\subsection*{Rebuilding protein conformational changes}

Like any motion, a conformational change of a protein, as for instance known by comparing structures A and B, can be described as a combination of motions along the normal modes of structure A, that is, with values of the corresponding normal coordinates obtained as follows:
\begin{equation}
q_k = \sum^{3N}_{i=1} a_{ik} \sqrt{m_i} \Delta r_i 
\label{eq:q}
\end{equation}
$\Delta r_i$ being the difference between the values of the i$^{th}$ coordinate of the protein for structures A and B, structure B being first fitted onto structure A so as to have $q_k = 0$ for the six zero-frequency modes associated to the overall translation and rotation of the protein. Note that, because the normal modes of a system form a basis set \cite{Wilson:55,Goldstein:56}, $I_k$, the percentage of involvement of mode $k$ in a given motion can be quantified as follows \cite{Ma:97}:   
\begin{equation}
I_k = 100 \frac{q_k^2}{\lVert \bf{\Delta r} \rVert}
\label{eq:involvement}
\end{equation}
with $\sum\limits^{3N} I_k =$ 100\%. Hereafter, it is assumed that mode $k$ is significantly involved in a conformational change if $I_k > 5$\%. 

Reciprocally, using eqn \ref{eq:dr} and the values of $n$ normal coordinates obtained through eqn \ref{eq:q}, the conformational change of a protein can be partially rebuilt, the rebuilding being perfect when $n = 3N$.

Hereafter, like in our previous study \cite{Mahajan:17}, the rebuilding of protein conformational changes is performed using their $n$ lowest-frequency modes. Note however that conformers thus obtained are then optimized with ROSETTA, as described in the previous section, ten independant optimizations being performed for each conformer built using eqn \ref{eq:dr}.

\subsection*{Elastic network models}

Protein normal modes were calculated with a standard Elastic Network Model \cite{Tirion:96,Hinsen:98,Bahar:00} (ENM), that is, a model where $V$, the (elastic) energy of the protein, is given by:
\begin{equation}
V = \frac{1}{2} k_{enm} \sum_{d_{ij}^0 < R_c} (d_{ij} - d_{ij}^0)^2
\label{eq:enm}
\end{equation} 
where $d_{ij}$ is the actual distance between atoms $i$ and $j$, $d_{ij}^0$ being their distance in the initial structure. $R_c$, a distance cutoff, is the only parameter of this ENM. Indeed, when, as assumed herein, $m_i = cste$, $k_{enm}$ is just an unit factor allowing for instance to specify the value of the lowest (non-zero) normal mode frequency of the system.   
 
Hereafter, two flavors of this kind of ENM are considered, namely, the popular C$_\alpha$-ENM \cite{Tama:01,Hinsen:98,Bahar:00}, which is built using only C$_\alpha$ atoms (with R$_c$ = 12 {\AA}), and the all-atom ENM noteworthy used by the Eln\'emo webserver \cite{Elnemo2} (with R$_c$ = 5 {\AA}), where atoms belonging to a given residue are assumed to behave like a rigid-body ---the so-called RTB approximation \cite{Tama:00,Durand:94,BNM,Grudinin:17}.   
 
In practice, normal mode calculations were performed with the enm\_modes software, which is freely available on HAL (\url{https://hal.archives-ouvertes.fr/hal-03142851}),
via the Software Heritage program \cite{Softwareheritage}.

\section*{Results}

\begin{figure}[t]
\includegraphics[width=8.0 cm]{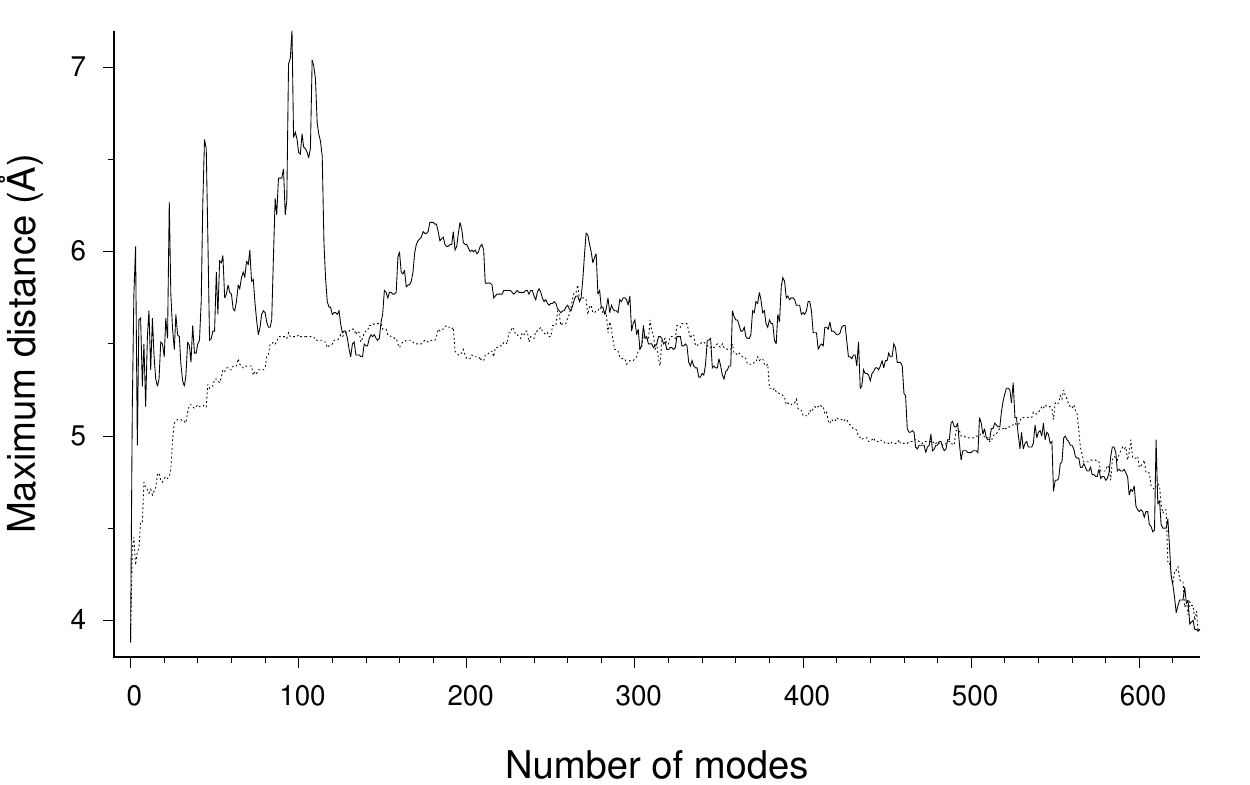}
\caption[]{
Maximum distance between consecutive C$_\alpha$ atoms along the sequence, as a function of the number of C$_\alpha$-ENM modes used to rebuild the conformational change of adenylate kinase.
Plain line: straight rebuilding. Dotted line: when the elastic energy of the conformer is minimized, keeping the normal coordinates used for the rebuilding fixed.  
}
\label{Fig:caca}
\end{figure}

\subsection*{The main bottleneck}

The main problem encountered when using only eqn \ref{eq:dr} for generating protein conformers is illustrated in Figure \ref{Fig:caca}, where the maximum distance between consecutive C$_\alpha$ atoms along the sequence, observed during the rebuilding of the conformational change of adenylate kinase, as known by comparing the open (PDB 4AKE \cite{4AKE}) and closed (PDB 1AKE \cite{1AKE}) forms, is shown.

With a C$_\alpha$-ENM of the open form of adenylate kinase, using eqn \ref{eq:dr} and the value of the lowest-frequency normal coordinate (eqn \ref{eq:q}) of the closed form, without any minimization of the elastic energy and no optimization by ROSETTA of the conformer thus obtained, the maximum distance between consecutive C$_\alpha$ atoms is 5.2 {\AA}, far from standard values, like the average one observed in the case of the open form, namely, 
3.8 $\pm$ 0.1 {\AA}. 

With the 102 lowest-frequency normal coordinates of the closed form, the maximum distance between consecutive C$_\alpha$ atoms of the conformer becomes as high as 7.2 {\AA}. However, its RMSd from the closed form is as low as 1.2 {\AA}, in line with previous works showing that the conformational change of adenylate kinase can be well described as a combination of motions along its low-frequency modes \cite{Tama:01,Tama:00}.     
So, while the overall conformational change of adenylate kinase is indeed very well described with the values of 102 normal coordinates, as illustrated in Figure \ref{Fig:caca}, chemical details look then quite ugly. As a matter of fact, more than $\approx$ 450 normal coordinates (70 \% of them) are needed in order not to have any distance between consecutive C$_\alpha$ atoms above 5 {\AA}.

On the other hand, Figure \ref{Fig:caca} shows that minimizing the energy of the elastic energy of the conformer thus obtained, while keeping the considered normal coordinates fixed, as done in our previous study \cite{Mahajan:17}, does not improve significantly the chemical accuracy of the conformers, distances between consecutive C$_\alpha$ atoms around 6 {\AA} being still observed. 

In our previous study \cite{Mahajan:17}, in order to cope with this issue, all conformers were further optimized with ROSETTA \cite{Rosetta:15}, using harmonic restraints on the protein backbone. Hereafter, as mentioned in the Methods section, optimized conformers are obtained so as to match a large set of distance restraints. Interestingly, with such an approach, the chemical accuracy of the generated conformers is guaranteed, since it only depends upon the quality of the initial structure, that is, the structure used for the normal mode analysis. 

While working with internal coordinates \cite{Go:82,Levitt:83,Chacon:14,Tirion:15} also secures the chemical accuracy of the generated conformers \cite{Go:85,Lavery:19}, note that variations of a backbone internal coordinate can yield steric clashes far away along the sequence, a problem which is the counterpart of the problem with chemical accuracy occurring when working with cartesian coordinates.  

\subsection*{Choice of the kind of ENM}

\begin{figure}[t]
\includegraphics[width=7.5 cm]{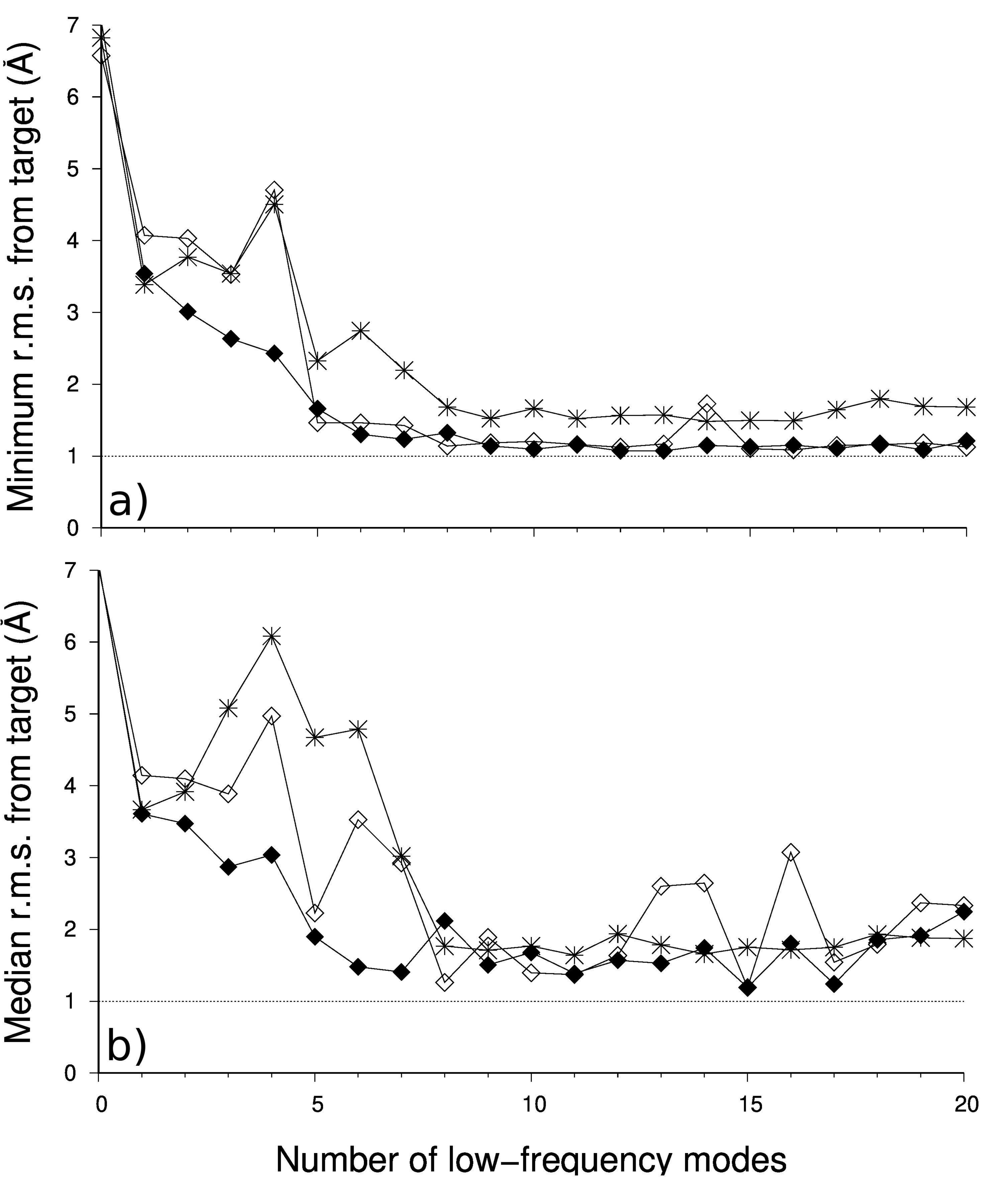}
\caption[]{
C$_\alpha$ root-mean-square deviation from adenylate kinase closed form (PDB 4AKE), as a function of the number of ENM modes used to rebuild the conformational change of adenylate kinase, starting from the open form (PDB 1AKE).
a) Minimum and b) median values obtained after ten ROSETTA optimizations, using modes obtained with a C$_\alpha$-ENM, with (open diamonds) or without (stars) minimization of the elastic energy, keeping the normal coordinates used for the rebuilding fixed; filled diamonds: using an all-atom ENM together with elastic energy minimization. Dotted line: corresponds to 1.0 {\AA}.
}
\label{Fig:adk}
\end{figure}

As shown in Figure \ref{Fig:adk}, in the case of the rebuilding of the conformational change of adenylate kinase, starting from the open form, using the values of less than 5--8 low-frequency normal coordinates of the closed form proves enough for obtaining optimized conformers less than 2 {\AA} away from the target. However, with the C$_\alpha$-ENM and no preliminary minimization of the elastic energy of the conformers, even with 20 low-frequency normal coordinates, no optimized conformer less than 1.5 {\AA} away is obtained, while otherwise conformers $\approx$ 1.0 {\AA} away from the closed form are found (Fig. \ref{Fig:adk}a).  

Noteworthy, with the all-atom ENM, a median RMSd from the closed form of 1.5 {\AA} is observed with as little as six low-frequency normal coordinates (Fig. \ref{Fig:adk}b). Moreover, with this model, the quality of the conformers obtained seems to depend less upon the exact number of low-frequency normal coordinates taken into account. This is the main reason why it was chosen for performing the rest of this work.

\subsection*{Rebuilding conformational changes}

\begin{figure}[t]
\hskip -0.2 cm
\includegraphics[width=8.0 cm]{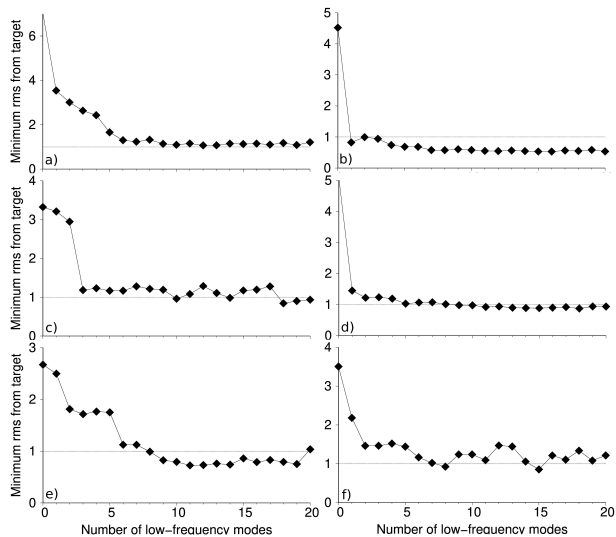}
\caption[]{
Minimum C$_\alpha$ root-mean-square deviation from the target, as a function of the number of all-atom ENM modes used to rebuild the conformational change of: a) Adenylate kinase (4AKE $\,\to\,$ 1AKE); b) LAO binding protein (2LAO $\,\to\,$ 1LST); c) Lysozyme T4 (177L $\,\to\,$ 178L); d) Glutamin binding protein (1GGG $\,\to\,$ 1WDN); e) NF-$\kappa$B (1LEI $\,\to\,$ 1RAM); f) Maltose binding protein (1OMP $\,\to\,$ 1ANF). Dotted line: corresponds to 1.0 {\AA}.}
\label{Fig:6cas}
\end{figure}

Figure \ref{Fig:6cas} shows the results obtained when rebuilding the conformational change of six proteins, including classic cases like adenylate kinase \cite{Tama:00,Tama:01,Doruker:16,Karplus:05}, lysozyme T4 \cite{Berendsen:98,Nicolay:06,Zacharias:08enm}, the lysine/arginine/ornithine (LAO) \cite{Tama:01,Chacon:14,Nicolay:06} or the maltose \cite{Tama:01,Tirion:96,Chacon:14} binding proteins.     

The more spectacular result is obtained with the LAO binding protein (Fig. \ref{Fig:6cas}b). Indeed, providing the value of a single normal coordinate, namely, the lowest-frequency one, proves enough for building, starting from the open form (PDB 2LAO \cite{oh:93}), a conformer less than 1 {\AA} away from the closed form (PDB 1LST \cite{oh:93}). However, such a success is not surprising since the percentage of involvement (eqn \ref{eq:involvement}) of the lowest-frequency normal mode of the open form in the conformational change of the LAO binding protein is very high ($I_7 =$ 86\%).

A result almost as convincing is obtained in the case of the glutamin binding protein (Fig. \ref{Fig:6cas}d), since providing the value of the lowest-frequency normal coordinate allows to build a conformer $\approx$ 1.5 {\AA} away from the closed form (PDB 1WDN \cite{1WDN}). Note that, in this case, the lowest-frequency mode is less involved in the conformational change ($I_7 =$ 76\%), with another one being also significantly, though marginally, involved, namely, the second lowest-frequency one ($I_8 =$ 5\%).  

While three of them are needed in the case of Lysozyme T4 (Fig. \ref{Fig:6cas}c), note that, in all six cases, conformers $\approx$ 1 {\AA} away from the target are obtained by specifying the values of no more than six low-frequency normal coordinates.

\begin{figure}[t!]
\includegraphics[width=8.0 cm]{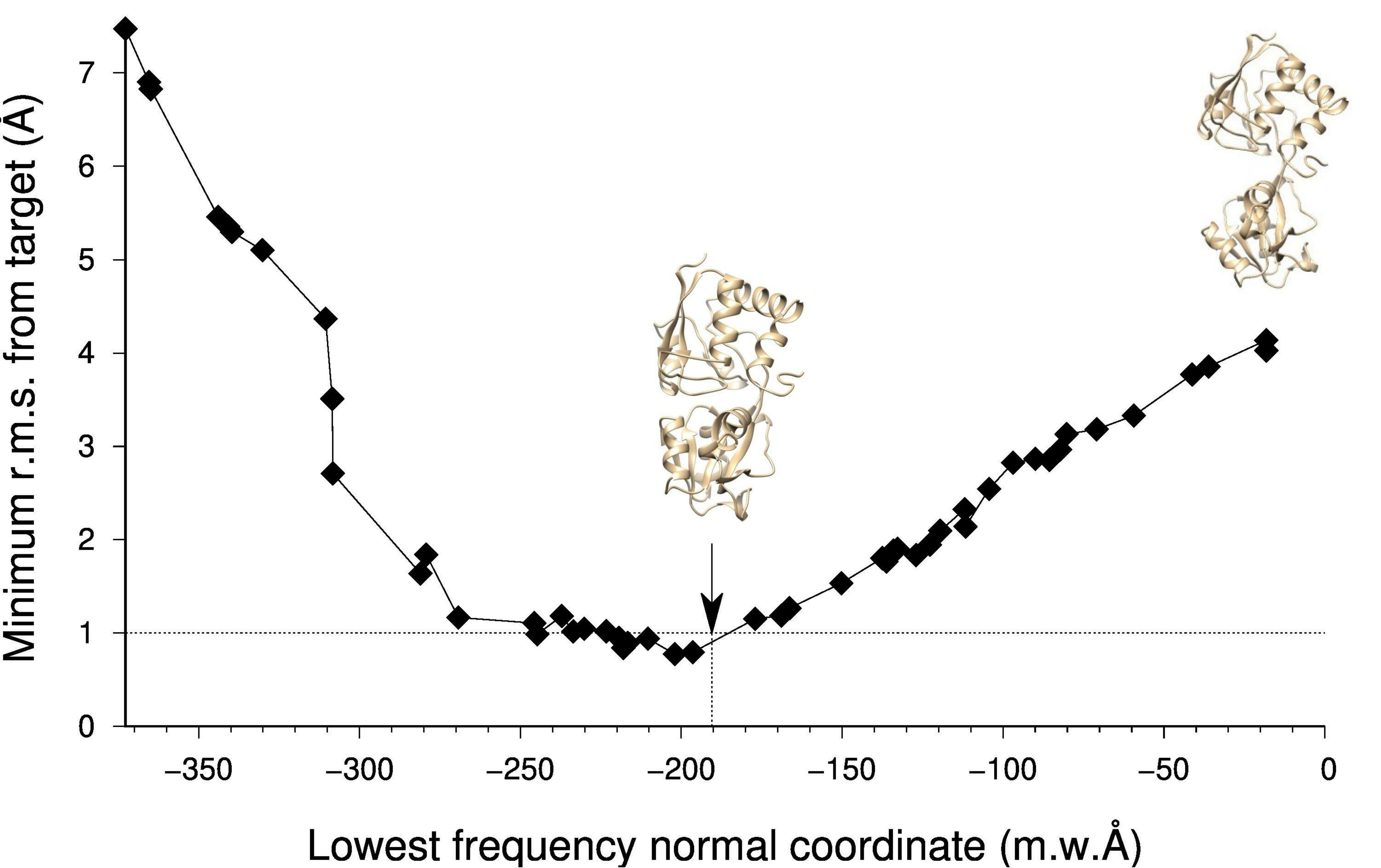}
\caption[]{
Minimum C$_\alpha$ root-mean-square deviation from the LAO binding protein closed form (PDB 1LST), as a function of the value of the lowest-frequency normal coordinate used to build the conformers, starting from the open form (PDB 2LAO). The arrow pinpoints the value observed for the closed form. Dotted line: corresponds to 1.0 {\AA}.
}
\label{Fig:laobp}
\end{figure}

\begin{table*}[]
\addtolength{\leftskip} {-0.2cm}
 \caption{Average and lowest energy of the conformers found close to the target (in kcal/mole per residue). Conformers were randomly generated using between one and five low-frequency modes of the initial ones.}
 \label{Table:confs} 
\begin{tabular}{|c|c|c|c|c|c|c|c|}
 \hline
 \multirow{2}{*}{Protein} &  \multirow{2}{*}{$\langle$Energy$\rangle^a$} & \multirow{2}{*}{Initial$\,\to\,$Target} & RMSd & Modes &  Number of   &  \multirow{2}{*}{$\langle$Energy$\rangle$} & Lowest \\
&  & &   ({\AA})  & used$^b$  &  conformers & & energy \\
 \hline
LAO bp           &	-3.27 $\pm$ 0.03 & 2LAO $\,\to\,$ 1LST & 3.4 & 7 & 38$^c$ & -3.2 $\pm$ 0.1 & -3.31 \\
Lysozyme T4      &  -2.78 $\pm$ 0.05 & 177L $\,\to\,$ 178L & 4.7 & 7--9 & 43$^d$ & -2.6 $\pm$ 0.2 & -2.84 \\ 
Adenylate kin. &  -3.02 $\pm$ 0.03 & 4AKE $\,\to\,$ 1AKE & 7.1 & 7--11 & 29$^e$ & -2.7 $\pm$ 0.2 & -2.94 \\
\hline
\end{tabular}

$^a$ Of the initial conformer, the average being over 100 relaxed conformations.\\
$^b$ Modes 1--6 correspond to overall translation and rotation motions.\\
$^c$ With a RMSd from target $\leq$ 1.0 {\AA}.\\
$^d$ With a RMSd from target $\leq$ 1.5 {\AA}.\\
$^e$ With a RMSd from target $\leq$ 2.0 {\AA}.
\end{table*}

In this respect, NF-$\kappa$B is an interesting case. Indeed, providing less than six values does not allow to build conformers significantly less than 2 {\AA} away from the target (PDB 1RAM \cite{NFKB}). Adding the sixth value means that the highest-frequency mode found significantly involved in the conformational change is taken into account. Doing so, conformers $\approx$ 1 {\AA} away from the target can be built (Fig. \ref{Fig:6cas}e). So, though the involvement of mode 12 may seem low ($I_{12}$ = 12\%), the information brought by this mode proves important for the accurate rebuilding of the conformational change of NF-$\kappa$B.  

\subsection*{Single mode exploration}

If the value of a single normal coordinate is enough for rebuilding the conformational change of a protein, it is then straightforward, starting from the initial conformer, to build conformers close to the target when the structure of the latter is unknown, for instance by randomly picking values for the relevant  
normal coordinate. 

This is illustrated in Figure \ref{Fig:laobp}, where 50 randomly chosen negative values of the lowest-frequency normal coordinate were considered, conformers being generated starting from the open form of the LAO binding protein. 
As above, for a given value of the normal coordinate, ten ROSETTA optimizations were performed.  
Among the 500 conformers thus obtained, 38 (8\%) were found less than 1 {\AA} away from the closed form. Moreover, compared to the energy of the open form, their average energy is similar, the lowest conformer energy observed being even lower than the average energy of the open form (see Table \ref{Table:confs}).   
Note that the values of the lowest-frequency normal coordinate used for generating such highly accurate conformers span a rather wide range, namely, between -245 and -196 mass-weighted {\AA} (Fig. \ref{Fig:laobp}). 

\subsection*{Subspace exploration}

Random exploration of a subspace with less than ten dimensions can also prove efficient, as illustrated by the two following examples, where the sampling was however limited by choosing the right sign for the normal coordinates, that is, the sign observed for the target, as well as maximum values twice as large as the value known for the target.

In the case of Lysozyme T4, such random sampling of the values of the three lowest-frequency normal coordinates of a mutant with a wild-type conformation (PDB 177L \cite{Matthews:90}) allowed to obtain 43 conformers (out of 1,000 trials) less than 1.5 {\AA} away from the target (PDB 178L \cite{Matthews:90}). Moreover, like in the case of the LAO binding protein, their average energy is rather low, the lowest conformer energy observed being also lower than the average energy of the initial one (see Table \ref{Table:confs}). 

In the case of adenylate kinase, random sampling of the values of the five lowest-frequency normal coordinates of the open form allowed to obtain 22 conformers (out of 5,000) less than 2.0 {\AA} away from the closed form.

\subsection*{Exploring with large jumps}

Using eqn \ref{eq:dr}, atomic displacements can be scaled, so as to generate conformers a given RMSd away from the initial one. 

In the case of adenylate kinase, conformers 6--7 {\AA} away from the open form (PDB 4AKE) were thus built, the scaling being performed after a random choice of the values of the five lowest-frequency normal coordinates. Note that, in this case, their sign was also chosen randomly.

First, 10,000 conformers were generated using eqn \ref{eq:dr}, and their elastic energy was minimized, keeping the five normal coordinates considered fixed. The 71 conformers found less than 3.5 {\AA} away from the closed form were then optimized with ROSETTA, as done above, allowing to obtain 7 other conformers less than 2.0 {\AA} away from the target. 

Though the average energy of the 29 conformers found close to the closed form of adenylate kinase is also low, at variance with both previous cases, the lowest conformer energy observed is significantly higher than the average energy of the open form (see Table \ref{Table:confs}).
This is not due to the energy gap expected between open and closed forms since, according to ROSETTA, in spite of the fact that ligands are missing, the closed form has the lowest average energy, namely, -3.17 $\pm$ 0.04 kcal/mole per residue.   
It may mean instead that 
the sampling performed was not extensive enough (the generated conformer the closest to the closed form being still 1.6 {\AA} away), or that
more than five normal coordinates should have been taken into account during the conformer generation process.

\begin{figure}[t!]
\includegraphics[width=7.5 cm]{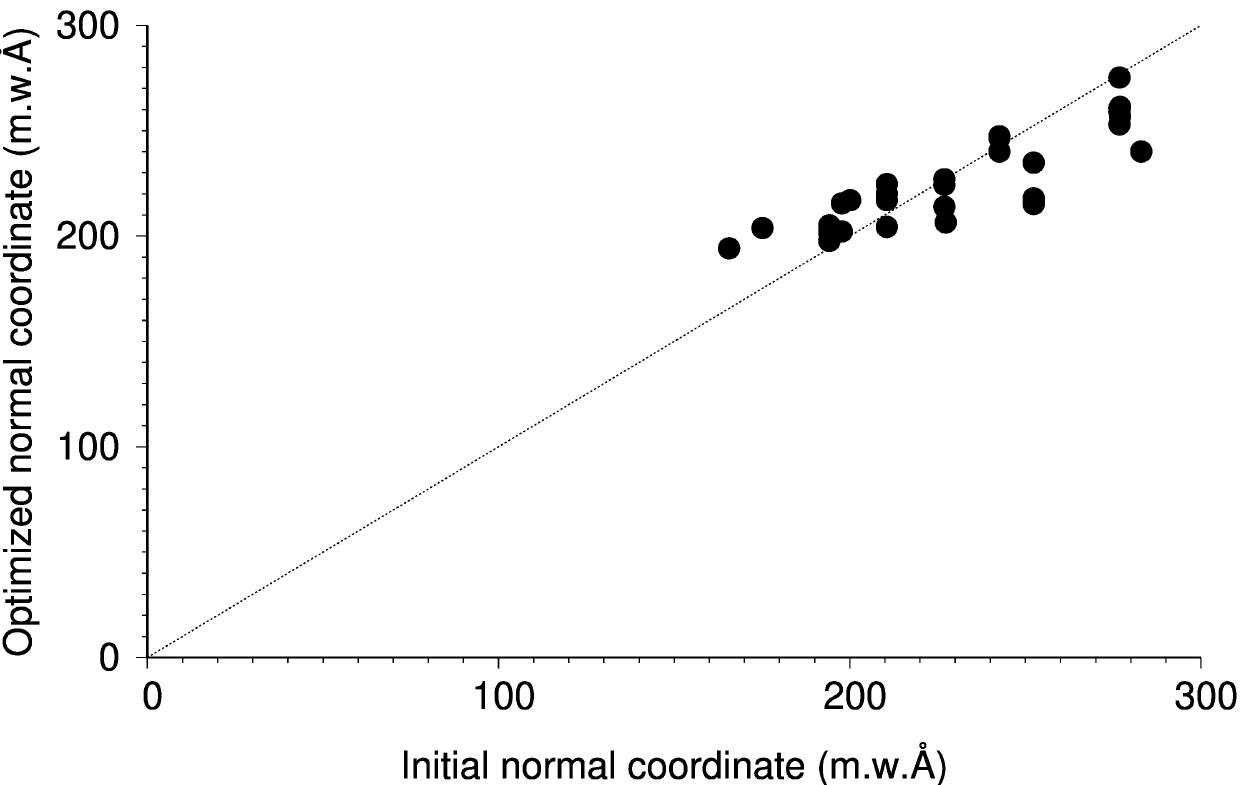}
\caption[]{
Optimized value of the lowest-frequency normal coordinate of adenylate kinase, as a function of its value before ROSETTA optimization, for conformers found less than 2.0 {\AA} away from the closed form, during a random exploration along the five lowest-frequency modes of the open form.
(0,0) corresponds to the open form.
The dotted line, which indicates equal values, is a guide for the eye.
}
\label{Fig:adkopt}
\end{figure}

\section*{The contribution of ROSETTA}

As illustrated in Figure \ref{Fig:adkopt} for the case of adenylate kinase, probably as a consequence of the large number of restraints taken into account, the value of the lowest-frequency normal coordinate does not change much upon optimization by ROSETTA. This suggests that the approach introduced herein should work with various optimization methods, noteworthy those developed specifically for solving the distance geometry problem  \cite{Wuthrich:87,Zwu:99,Malliavin:19}.  

\section*{Conclusion}

The approach proposed in the present study allowed to rebuild accurately (RMSd $\approx$ 1 {\AA}) six large amplitude (RMSd $>$ 3 {\AA}) conformational changes, using the values of at most six low-frequency normal coordinates (Fig. \ref{Fig:6cas}). 

This result noteworthy means that, in a predictive context, as a consequence of the low-dimensionality of the corresponding subspace, conformers close to the target can be obtained in a straightforward way, for instance by random sampling, as done herein for three classic cases, namely, the conformational changes of the LAO binding protein (Fig. \ref{Fig:laobp}), lysozyme T4 and adenylate kinase (Table \ref{Table:confs}).  

More efficient algorithms should significantly speed up the conformer generation process. Note however that it is already a fairly quick one, generating an optimized conformer taking a few minutes \textit{only}, on a single standard CPU.

So, a timely task left is to delineate the field of application of this approach. In the present study, a handful of proteins with clear-cut large amplitude domain motions were scrutinized. 
Proteins experiencing motions more arbitrary or of a more subtle nature need now to be considered. 

\section*{Acknowledgments}
This study is a sequel of the work undertaken with Dr Swapnil Mahajan during the BIP:BIP \textit{Projet d'Avenir} (grant ANR-10-BINF-03-04).

\end{document}